# Taxonomy of AISecOps Threat Modeling for Cloud Based Medical Chatbots

Ruby Annette J, Aisha Banu *, Sharon Priya S *, Subash Chandran

*Senior Member IEEE, *BSA Crescent Institute of Science and Technology, NEC Corporation of America*

## Abstract

Artificial Intelligence (AI) is playing a vital role in all aspects of technology including cyber security. Application of Conversational AI like the chatbots are also becoming very popular in the medical field to provide timely and immediate medical assistance to patients in need. As medical chatbots deal with a lot of sensitive information, the security of these chatbots is crucial. To secure the confidentiality, integrity, and availability of cloud-hosted assets like these, medical chatbots can be monitored using AISecOps (Artificial Intelligence for Secure IT Operations). AISecOPs is an emerging field that integrates three different but interrelated domains like the IT operation, AI, and security as one domain, where the expertise from all these three domains are used cohesively to secure the cyber assets. It considers cloud operations and security in a holistic framework to collect the metrics required to assess the security threats and train the AI models to take immediate actions. This work is focused on applying the STRIDE threat modeling framework to model the possible threats involved in each component of the chatbot to enable the automatic threat detection using the AISecOps techniques. This threat modeling framework is tailored to the medical chatbots that involves sensitive data sharing but could also be applied for chatbots used in other sectors like the financial services, public sector, and government sectors that are concerned with security and compliance.

## 1. Introduction

Medical chatbots are becoming popular these days as they could be used very effectively to provide timely help to the patients who are in remote or unreachable areas. The medical bots are also usually hosted in the cloud to adhere to the dynamic scaling of the data and the number of users [1]. The conversational AI involved in medical chatbots may frequently encounter many sensitive information about the patient like their personally identifiable information, social security numbers, their medical records etc [2]. Because of the high exposure of medical chatbots to the personal information of the patients, the fraudsters have targeted this industry as a major target. These attackers seek to steal personally identifiable information (PII) and other highly classified patient information, posing challenges for the chatbot's security architecture [3]. Cloud computing has been widely used to scale up the resources on demand in various fields like finance, medical[4] etc. to enable scaling up of resources on demand. Since the applications can be scheduled [5] to various nodes in the cloud environment, the applications can perform their tasks efficiently by scaling up or down the number of compute resources [6]. Monitoring and ranking the cloud services [7] based on the multiple functional and nonfunctional criteria requirements for selecting the right cloud services [8] to ensure the security of the computing assets hosted in the cloud as well as in the non-cloud environment is very crucial in the field of medicine as sensitive personal and medical data are stored in these digital assets. AISecOPs is a part of AIOps [9] an emerging field that integrates three different but interrelated domains like the IT operation, AI, and security as one domain, where the expertise from all these three domains are used cohesively to secure the cyber assets [10]. AISecOps (Artificial Intelligence for Secure IT Operations) [11] is an implementation and delivery focused technique that combines full-stack telemetry data and automation to assist enterprises in delivering a dependable, secure, and cost-effective IT service that is continuously optimized for performance and security. AISecOps is a term coined by Gartner to describe the use of AI capabilities like natural language processing (NLP) and machine learning models to automate and streamline operational activities to ensure security. AISecOps makes use of big data, analytics, and machine learning to a) Collect and consolidate the massive and ever-increasing volumes of data produced by various IT infrastructure components, application requests, performance-monitoring tools that help in monitoring the security of the systems. b) Identify critical events and patterns linked to security concerns by intelligently shifting 'signals' out of the 'noise.' c) Diagnose fundamental causes and communicate them to security team for quick reaction and remediation or in some situations, automatically resolve issues without the need for human participation. d)AISecOps helps IT operations teams to respond more quickly and even proactively to potential security threats by merging numerous distinct, manual IT operations technologies into a unified, intelligent, and automated IT operations platform with end-to-end visibility and context. e) It bridges the gap between a more diversified, dynamic, and difficult-to-monitor IT landscape and siloed teams on the one hand, and user expectations for application performance and availability with little or no downtime on the other. Most experts believe that AIOps is the way of the future for IT operations management, and demand is only growing as businesses focus more on digital transformation efforts. This work maps the popular STRIDE framework for cyber security threat modelling to the various components of the conversation AI based medical chatbots architecture. STRIDE is a threat modeling methodology created and published by Microsoft employees in 1999 [12]. The STRIDE threat model examines the probable consequences of several threats to a system. STRIDE is the acronym for the different types of important threats to be identified that includes Spoofing(S), Tampering(T), Repudiation(R), Publication of information (P), Service interruption (I) and Privileges Elevation (E). STRIDE makes it feasible to uncover potential attack vectors for the conversational AI chatbots. This helps to decide on the key metrics to be collected for the chatbot security analysis. The AISecOps techniques can then be applied to analyze the danger and impact of any potential threat using this knowledge and devise actions to mitigate it. Ye W et al [13] have explained the various cyber threats that are associated with the different components of the chatbots, but their work is not focused on threat modelling and mapping the threats to the popular STRIDE framework for threat modelling.

This work discusses on the following topics: a) A detailed description of the components of a cloud-based AI conversational Medical Chatbots is given in section II b) The STRIDE threat modeling for medical chatbot components is explained in section III c) The proposed taxonomy of cyber security threat modelling for cloud based medical chatbots is explained in detail in section IV d) Section V provides a detailed discussion and conclusion.

## 2. Components of Cloud-based AI conversational Medical Chatbots

Chatbots have become increasingly popular in a variety of applications, including recommender systems that recommend a variety of products and services from clothing to recommending highly technical services like cloud renderfarm services for medical images rendering [14]. They allow us to carry out activities that are activated by both written and spoken verbal orders and could be very useful in the field of medicine as well. The medical chatbots could be used very effectively to provide timely help to the patients who are in remote or unreachable areas. The medical bots are also usually hosted in cloud that are carefully chosen based on the functional and nonfunctional requirements of the overall system to adhere to the dynamic scaling of the data and the number of users. The conversational AI involved in medical chatbots may frequently encounter many sensitive information about the patient like their personally identifiable information, social security numbers, their medical records etc. [15]. Because of the high exposure of medical chatbots to the personal information of the patients, the fraudsters have targeted this industry as a major target. These attackers seek to steal personally identifiable information (PII) and other highly classified patient information, producing challenges for the chatbot's security architecture [16].

The components or modules that are commonly available in chatbots include the following modules. Client Module is the first part of the chatbot with which the patient interacts with in a medical chatbot, it also includes the other applications that the medical chatbot may operate or may get connected to in the process of chatting with the patient. The second module is the communication module that includes all the components of the infrastructure that is required for sending the user / patient messages from the client / patient module to the response generation module and then from the response generation module to the database module. The next one is the response module that deals with all the tasks involved in responding back to the user / patient with the relevant information. Hence, this layer comprises of all the Natural Language Processing (NLP) and or the

Natural Language Understanding NLU and or the question-and-answer generation module etc. The module that translates the incoming textual command into an intent, that communicates the patient's intents, if the frontend is not used. The purpose of the phrase "When is the doctor's appointment for scan today?" is to learn about the appointment details of the patient, and the processing engine should be properly trained to recognize similar-sounding words with the same semantic meaning. The NLP (Natural Language Processing) toolset on the other hand includes several services. These are often based on neural networks that learn from instances of specific words used in a variety of contexts and purposes. Thus, this module is the core module that involves a lot of scope for Artificial Intelligence using machine learning models to understand the user's / patient's input message and create an appropriate response according to the input / request. The final module is the database module, and it is the module where all the information about a conversation is stored which may include message history etc. It may also contain the knowledge base in the form of knowledge graphs to reason and understand the about the environment around them, some chatbots employ the knowledge graph. These knowledge graphs may be helpful in the normal chatbots that does not use Natural Language Understanding NLU and or the question-and-answer generation techniques.

## 3. STRIDE Threat Modeling for Medical Chatbot Hosted in Cloud

Since the medical chatbot application are usually used by many patients scattered in various geographical locations, to ensure availability and data privacy, they need to be scheduled to various nodes in the cloud environment. Scheduling these applications on cloud will enable them to perform their tasks efficiently by scaling up or down the number of compute resources [17]. Cloud computing has been widely used to scale up the resources on demand in various fields like finance, medical images rendering etc. to enable scaling up of resources on demand. Monitoring and ranking the cloud services [18] based on the multiple functional and non-functional criteria requirements for selecting the right cloud services [19] to ensure the security of the computing assets hosted in the cloud as well as in the non-cloud environment is very crucial in the field of medicine as sensitive personal and medical data are stored in these digital assets. AISecOPs is an emerging field that integrates three different but interrelated domains like the IT operation, AI, and security as one domain, where the expertise from all these three domains are used cohesively to secure the cyber assets. Threat modeling is an exercise that is used to assess potential risks and attack routes for a system. It is possible to conduct risk analysis and build counter measures and strategies to manage and reduce these risks using this knowledge. Using a threat modeling framework gives the threat modeling process structure and may contain additional benefits like proposed detection tactics and counter measures to be taken.

This work maps this popular STRIDE framework for cyber security threat modeling to the various components of the conversation AI based medical chatbots architecture. STRIDE is a threat modeling methodology created and published by Microsoft employees in 1999. The STRIDE threat model examines the probable consequences of several threats to a system. STRIDE is the acronym for the different types of important threats to be identified that includes Spoofing(S), Tampering(T), Repudiation(R), Publication of information (P), Service interruption (I) and Privileges Elevation (E). STRIDE makes it feasible to uncover potential attack vectors for the conversational AI chatbots. This helps to decide on the key metrics to be collected for the chatbot security analysis. The AISecOps techniques can then be applied to analyze the danger and impact of any potential threat using this knowledge and devise actions to mitigate it. By examining these potential effects or goals and determining how they can be achieved, it is possible to find attack routes for the medical chatbot under examination. Using this knowledge, it is feasible to assess the hazard and impact of any prospective threat and design mitigation strategies.

The strategy is to break down the system into key components, assess each one for hazard susceptibility, and then use the STRIDE method to reduce the threats. The process is then continued until there are no more risks to be concerned about. When we have succeeded splitting the system down into components and mitigate all threats to each one, we may claim that it is secure. Modeling cyber security vulnerabilities that are feasible at each component level of AI conversational chatbots is essential. It aids in identifying the metrics that need to be watched to employ AISecOps methodologies to discover aberrant patterns in talks and chatbot connection networks.

# 4. Taxonomy of STRIDE Threat Modeling for Cloud Based Medical Chatbots

A taxonomy of the cyber security threat modeling for cloud based medical chatbots using the STRIDE method is proposed in this work. The taxonomy maps the STRIDE categorized cyber threats to the various components of the medical chatbots. This way of creating the taxonomy helps to model the cyber threats that needs to monitor and captured to ensure security of the chatbots and identify the potential threats and the responses to be ready with in case of a security violation. The categories of the treats based on the STRIDE that identifies the different types of important threats to be identified that includes Spoofing(S), Tampering(T), Repudiation(R), Publication of information (P), Service interruption (I) and Privileges Elevation (E) is mapped to each component of the medical chatbot as given in the Figure 1.

## 4.1. Spoofing (S):

Spoofing (S) is a type of fraud in which someone or something impersonates a legitimate source, business, colleague, or other trusted contact to collect personal information, obtain money, spread malware, or steal data by forging the sender's identity. Spoofing is the most common attack related with the chatbot's Client module [20]. The following types of spoofing are possible in the chatbots' Client module, namely the a) IP Address Spoofing, b) ARP Spoofing c) Man-in-the-middle attack.

### 4.1.1. IP Address Spoofing

It is a network protocol for sending and receiving messages over the internet. Every email message sent includes the sender's IP address in the message header (source address). Hackers and scammers alter the header data by changing the source address to mask their identity. emails appear to have originated from a trustworthy source.

### 4.1.2. ARP Spoofing

ARP spoofing is a hacking technique that redirects network traffic to a hacker. Spoofing is the process of sniffing out LAN addresses on both wired and wireless LAN networks. The goal of this type of spoofing is to send fake ARP communications to Ethernet LANs, causing traffic to be changed or completely stopped. ARP's primary function is to match the IP address to the MAC address. Spoofed messages will be sent across the local network by attackers. The user's MAC address will be mapped to his IP address in this answer. As a result, the attacker will have complete access to the victim's computer.

### 4.1.3. Man-in-the-Middle Attacks

As the name implies, communication between the original originator of the message and the intended receiver is intercepted. The substance of the message is then altered without the knowledge of either party. In the packet, the attacker adds his own message [21].

## 4.2. Tampering Data (T) / Data Tampering (T)

The act of purposely modifying, destroying, manipulating, or editing data through unauthorized channels is known as data tampering. There are two states of data: in transit and at rest. Data could be intercepted and tampered with in both cases. Data transfer is at the heart of digital communications. When data packets are sent unencrypted, for example, a hacker can intercept the data packet, change its contents, and change its destination address. A system application can suffer a security breach if data is stored at rest, and an unauthorized intruder could use malicious code to alter the data or underlying computer code. The intrusion is hostile in both cases, and the consequences for the data are always disastrous. It's one of the most serious security risks that any app, program, or company can face. In both the Response generation and database modules, data manipulation is feasible [22].

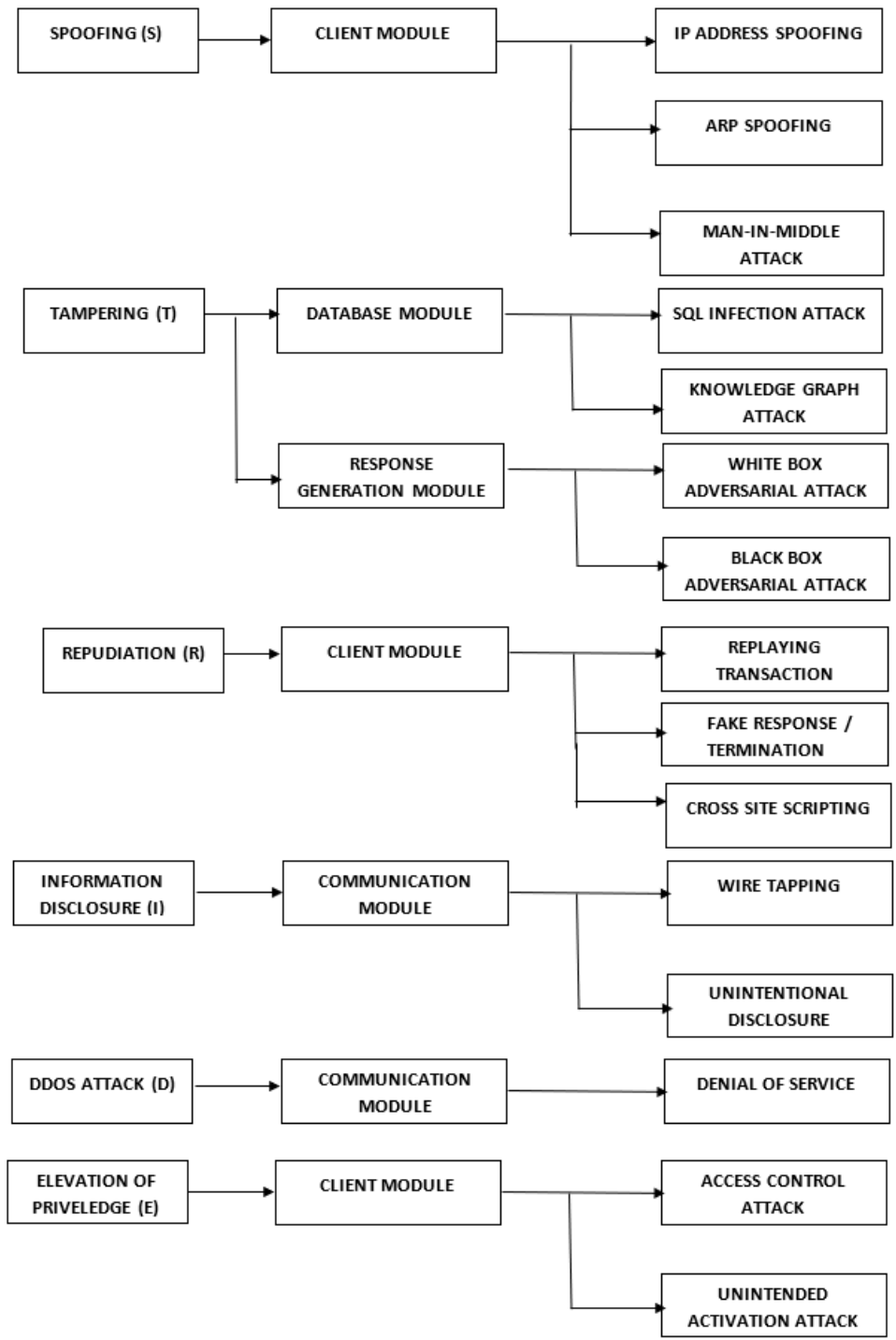

**Figure 1.** Taxonomy of STRIDE Threat Modeling for Cloud Based Medical Chatbots

*4.2.1. Data tampering attacks in chatbot database modules*

    Any relevant information to the conversation can be looked up in the database module of the chatbot. It can, for example, query a knowledge graph to produce a well-informed response. Attacks against the database module could jeopardize the privacy of millions of users and drastically alter the chatbot's behavior. The two

important attacks under this category is the SQL injection attack and the attack on the knowledge graph database.

SQL injection attacks are a common issue in programs that use SQL as a datastore. These attacks use carefully constructed inputs to force the database to undertake unwanted activities such as data modification or the return of sensitive data. Data injections can happen through server variables or even cookies, according to Halfond et al. [23]. An attacker may, for example, conceal an injection attack in a server variable that is activated if the database is told to record the contents of that variable. A variety of solutions can be used to combat SQL injection attacks. The main source of this vulnerability, however, is a lack of input validation. As a result, any effective defense against these attacks necessitates the developer spending adequate time cleaning and validating data [24].

Attacks on Knowledge Graphs used by some chatbots to reason and understand the about the environment around them is also very common and needs to be monitored. Knowledge graphs are a particular type of database that captures the relationships between various entities that are interconnected with each other and can be queried using special type of query language designed for graphs to extract the relationship between the entities and hence identify the impact or dependency of each component on the other. To create a knowledge graph usually a taxonomy [25] of the entities is created by identifying the important key role players in the environment and then the taxonomy is used to create an ontology that captures the different relationships between those entities and the other ways those entities may be represented in the knowledge graph. A social network graph is the most basic example of a knowledge graph, which represents the relationships between real-world items. Hence any attack that modifies the relationship between the entities will cause a huge impact on the inference made using the knowledge graph [26] Chatbots that are used for recommending items or generates recommendations based on the inputs given by the user make use the concepts of taxonomy [27], ontologies and knowledge graphs in their underlying architecture to overcome the cold start problem in recommending appropriate items to the users.

*4.2.2. Attacks to tamper with data in the chatbot's Response Generation module:*

Samples of Adversarial Voices (E) To work successfully, many personal assistants rely on a voice recognition component incorporated in the client module. These voice recognition features, however, are not always accurate. Hackers have been known to dupe personal assistants into carrying out their orders in the past. One can create hostile voice samples to trick the voice recognition module contained in the client program. White box attacks and black box attacks are the two types of attacks that can be used.

White box attacks [28] make use of model information, whereas black box attacks [29] make no assumptions about the model. The basic purpose in each case is to mask the antagonistic voice command that the attacker is attempting to elicit. One technique to do this is to introduce a disturbance to the original speech sample that causes the voice recognition module to misinterpret it but is undetectable to humans. Voice commands may be embedded in seemingly benign audio sources such as music in more complex attacks. The authors tweaked the innocent tune just enough for the voice recognition module to pick up on the antagonistic voice commands, but not humans. In the voice domain, countermeasures against hostile samples have been presented. White box attacks are simple to thwart if the model is kept hidden. However, adversarial samples can be created even in black box situations, thus relying too heavily on the model's secrecy is a bad idea. Other strategies aim to increase the amount of effort required to carry out a successful attack. The assault may become impossible if the adversary must query the model multiple times. The model can also be retrained by the developers. Mistakes in the voice recognition module can be corrected via adversarial training.

## 4.3. Repudiation

Repudiation is the inability to prove the existence of a transaction between two parties. When executing session layer assaults, the hacker's purpose is to perform repudiation. Nonrepudiation, on the other side, is having absolute confirmation of the parties' identities in a completed transaction. Certain sorts of transactions, for example, necessitate a nonrepudiation procedure. Signing electronic documents, exchanging money electronically, and purchasing a goods online with a credit card, for example, all require a nonrepudiation mechanism to be legally binding. When consumers access web material, hackers frequently utilize a repudiation attack. To carry out their repudiation assault, hackers can use Java or ActiveX scripts, port-scanning programs, masquerading, and eavesdropping on the chatbot's client module. Insider hackers can also perform a repudiation attack on the chatbot's Response Generation module by replacing the original verified NLP model with a backdoored NLP model that allows for additional attacks [30].

*4.3.1. Replaying Transactions*

In chatbots, replaying transactions is a sort of repudiation attack in which an attacker captures an earlier transaction and submits it later, triggering a replay attack. The majority of privacy dangers stem from a user's lack of privacy control. The lack of privacy control by user leads to leads to other types of issues like Lack of Consent, where a transaction is being carried out without the consent of a user. The other associated issue is the lack of control and transparency, where the users have little control on the way transaction is being carried out.

*4.3.2. Fake response / Termination*

Chatbots make be compromised by an attack to provide fake response which may include wrong or misleading information to the users. This is a fatal attack in case of medical chatbots, where the misleading information to the patients could cause a lot of damage to the decisions made by the patient. Hence, it is important to ensure that the chatbot provides the correct information. A strategy that may be used for fake termination is to impersonate termination with a response like "goodbye." but can continue recording the user even while the user thinks the app is closed. However, without an audio input, the chatbots are usually programmed to forcefully terminate a chat application. The method of including a silent audio file in their attack to feed to the chatbot to get around this issue has also been discussed [31]. By creating a blacklist of questionable chatbot responses, developers can design counter measures to these attacks.

*4.3.3. Cross Site Scripting (XSS)*

The XSS vulnerability is exploited by entering text in the chatbot frontend that includes malicious JavaScript code and then running the injected code. To take advantage of an XSS flaw, the attacker must persuade the victim to enter malicious input content. The attacker convinces the victim to click on a link that leads to the chatbot's frontend and contains malicious code. The webpage has harmful code inserted into it. Without the victim's knowledge, it reads the victim's cookies and sends them to the attacker. These cookies could be used by the attacker to get access to the victim's account on the company's website.

**4.4. Information Disclosure (I)**

The most important concern with chatbots is securing the information that the user submits during a discussion and guaranteeing that no third party may access, read, or exploit it in any manner. Information supplied through a chat might be vulnerable to not only financial or identity fraud, but also to hazards of a more sensitive kind that go beyond collecting and selling user data. Emotional engagement in the conversation and the AI's level of empathy are two aspects that may inspire people to divulge more personal information, such as health information, sexual orientation, and routine conditions.

**4.5. DDoS (Distributed Denial of Service) attack**

In the scenario of a DoS attack, the sender and recipient's message packets are intercepted, and the source address is faked, in this technique. The link has been snatched. As a result, the recipient is inundated with packets that exceed their bandwidth or capabilities. This causes the victim's system to become overloaded, effectively shutting it down. DDoS attacks [32] are staged using several systems, whereas DoS attacks are set up using a single system that gains access to the bank's servers and networks. As a result, detecting and blocking it is extremely tough. Hackers can acquire access to the organization's computers and network by simply exposing the service Chabot to malware. After getting access to the system, cyber criminals collect all personally identifiable information (PII) and sensitive data for nefarious reasons. A DDoS attack causes the server that supports customer and conversational AI to be overwhelmed with requests, causing the system to break.

A chatbot is also under the DoS attack when an attacker makes it hard for the Chatbot to give service and react to the customer's queries. DDoS assaults (distributed denial of service attacks) are widespread in the Communication layer of the chatbot, where the access to servers, devices, networks, and applications is restricted by hackers.

### 4.6. Privilege Elevation (E)

A privilege elevation attack is a sort of network intrusion that uses programming faults or design defects in the chatbots to provide the attacker increased access to the network and its data and applications. Not every system hack will grant an unauthorized user complete access to the targeted system right away.

Zhang et al [33] for example, discovered showed nearly half of the 156 users in their survey attempted to move from one ability to another in the middle of a conversation. An example of some actual privilege elevation attack due to design fault in the chatbot may be when assuming that the user has downloaded a malicious skill masquerading as an online hotel reservation system. This harmful skill could only collect the user's trip information without generating suspicion on its own. However, once the user selects the malicious software that recommends gyms nearby or medical doctors nearby, in this case the chatbot is now free to gather the user's medical information.

## 5. Discussion and Conclusion:

By communicating with users in a human-like manner, a medical chatbot aids the job of a healthcare provider and helps them enhance their performance. These intelligent medical chatbots could aid physicians, nurses, therapists, patients, and their families in a variety of ways. They can step in and cut down on the time they spend on things like connecting people and organizations with first responders, giving health-related information to users, direction for patient medication management and answer the Frequently Asked Questions (FAQs). However, it is crucial to remember that the medical chatbots should be designed carefully with the potential STRIDE threat modelling to build a medical chatbot service that can endure system faults and cyber-attacks while maintaining user privacy by taking steps to protect their personal information and create trustworthiness. This proposed taxonomy of cyber security threat modelling for cloud based medical chatbots would help to model the cyber threats that needs to monitor and captured to ensure security of the chatbots and identify the potential threats and respond immediately in case of a security violation.

## References

[1] Duvvuri V, Guan Q, Daddala S, Harris M, Kaushik S. Predicting Depression Symptoms from Discord Chat Messaging Using AI Medical Chatbots. In 2022 The 6th International Conference on Machine Learning and Soft Computing 2022 Jan 15 (pp. 111-119).

[2] Walss M, Anzengruber F, Arafa A, Djamei V, Navarini AA. Implementing medical chatbots: An application on hidradenitis suppurativa. Dermatology. 2021;237(5):712-8.

[3] Safi Z, Abd-Alrazaq A, Khalifa M, Househ M. Technical aspects of developing chatbots for medical applications: scoping review. Journal of medical Internet research. 2020 Dec 18;22(12): e19127.

[4] Chung K, Park RC. Chatbot-based healthcare service with a knowledge base for cloud computing. Cluster Computing. 2019 Jan;22(1):1925- 37.

[5] Ouerhani N, Maalel A, Ben Ghézela H. SPeCECA: a smart pervasive chatbot for emergency case assistance based on cloud computing. Cluster Computing. 2020 Dec;23(4):2471-82.

[6] Annette, J. Ruby, Aisha Banu, and P. Subash Chandran. "Comparison of Multi Criteria Decision Making Algorithms for Ranking Cloud Renderfarm Services. Indian Journal of Science and Technology Vol. 9, (2016).

[7] Annette J, Ruby, Aisha Banu W. "Ranking and Selection of Cloud Renderfarm Services", Sadhana, 44:7, 2019.

[8] Annette J, Ruby, Aisha Banu W, Subash Chandran" Classification and Comparison of Cloud Renderfarm Services for Recommender Systems". Lecture Notes on Data Engineering and Communications Technologies, Springer, 2019.

[9] Müller HA, Rivera LF, Jiménez M, Villegas NM, Tamura G, Akkiraju R, Watts I, Erpenbach E. Proactive AIOps through digital twins. in Proceedings of the 31st Annual International Conference on Computer Science and Software Engineering 2021 Nov 22 (pp. 275-276).

[10] An L, Tu AJ, Liu X, Akkiraju R. Real-time Statistical Log Anomaly Detection with Continuous AIOps Learning. In CLOSER 2022 (pp. 223- 230).

[11] Runzi ZH, Wenmao LI. An Intelligent Security Operation Technology System Framework AISecOps. Frontiers of Data and Computing. 2021 Jul 9;3(3):32-47.

[12] Threats - Microsoft Threat Modeling Tool - Azure | Microsoft Docs (last browsed on : 12th May 2023)

[13] Ye W, Li Q. Chatbot Security and Privacy in the Age of Personal Assistants. In 2020 IEEE/ACM Symposium on Edge Computing 2020 Nov 12 (pp. 388-393). IEEE.

[14] Annette J, Ruby, Aisha Banu W, Chandran Subash. A Multi Criteria Recommendation Engine Model for Cloud Renderfarm Services." International Journal of Electrical and Computer Engineering, Vol.8, 2018.

[15] Rathnayaka P, Mills N, Burnett D, De Silva D, Alahakoon D, Gray R. A Mental Health Chatbot with Cognitive Skills for Personalised Behavioural Activation and Remote Health Monitoring. Sensors. 2022 Jan;22(10):3653.

[16] Emily Dinan, Stephen Roller, Kurt Shuster, Angela Fan, Michael Auli, and Jason Weston. Wizard of Wikipedia: Knowledge-powered conversational agents. In Proceedings of the International Conference on Learning Representations. 2019.

[17] Annette J, Ruby, Aisha Banu W, Chandran Subash. "RenderSelect: A Cloud Broker framework for cloud renderfarm services selection." International Journal of Applied Engineering Research, Vol.10, No.20 (2015).

[18] K Niha, WA Banu, R Annette. "A Cloud Service Providers Ranking System Using Ontology" International Journal of Scientific & Engineering Research, 6 (4), 41-45, 2015.

[19] Annette, Ruby; Banu. W, Aisha. "A Service Broker Model for Cloud based Render Farm Selection." International Journal of Computer Applications, vol. 96, issue 24, pp. 11-14, 2015.

[20] Osanaiye OA. Short Paper: IP spoofing detection for preventing DDoS attack in Cloud Computing. In 2015 18th International conference on intelligence in next generation networks, 2015 Feb 17 (pp. 139-141). IEEE.

[21] T. Lauinger, V. Pankakoski, D. Balzarotti, and E. Kirda, "Honeybot, your man in the middle for automated social engineering." in USENIX, 27 April 2010.

[22] Alberti M, Pondenkandath V, Wursch M, Bouillon M, Seuret M, Ingold R, Liwicki M. Are you tampering with my data? in Proceedings of the European Conference on Computer Vision Workshop 2018 (pp. 0-0).

[23] W. G. Halfond, J. Viegas, A. Orso ., "A classification of SQL-injection attacks and countermeasures," in Proceedings of the IEEE International symposium on secure software engineering, vol. 1. IEEE, 2006, pp. 13–15

[24] Y.-W. Huang, S.-K. Huang, T.-P. Lin, and C.-H. Tsai, "Web application security assessment by fault injection and behavior monitoring," in Proceedings of the 12th international conference on World Wide Web, 2003, pp. 148–159.

[25] Annette, J. Ruby, W. Aisha Banu, and P. Subash Chandran. "Rendering-as-a-Service: Taxonomy and Comparison". Procedia Computer Science 50 (2015): 276-281, Elsevier.

[26] H. Zhang, T. Zheng, J. Gao, C. Miao, L. Su, Y. Li, and K. Ren, "Data poisoning attack against knowledge graph embedding," in Proceedings of the Twenty-Eighth International Joint Conference on Artificial Intelligence, 2019.

[27] Annette J, Ruby, Aisha Banu W, and Shriram. "A Taxonomy and Survey of Scheduling Algorithms in Cloud: Based on task dependency." International Journal of Computer Applications, 82.15 (2013): 20-26.

[28] Wei Emma Zhang, Quan Z. Sheng, and Ahoud Abdulrahmn F. Alhazmi. Generating textual adversarial examples for deep learning models: A survey. CoRR, abs/1901.06796, 2019.

[29] Mohit Iyyer, John Wieting, Kevin Gimpel, and Luke Zettlemoyer. Adversarial example generation with syntactically controlled paraphrase networks. In Proceedings of the 2018 Conference of the North American Chapter of the Association for Computational Linguistics: Human Language Technologies, NAACL-HLT 2018, New Orleans, Louisiana, USA, June 1-6, 2018, Volume 1 (Long Papers), pages 1875–1885, 2018.

[30] P. Neekhara, S. Hussain, S. Dubnov, and F. Koushanfar, "Adversarial reprogramming of text classification neural networks," in ACL, 2019.

[31] H. Zhong, C. Xiao, C. Tu, T. Zhang, Z. Liu, and M. Sun, "How does NLP benefit legal system: A summary of legal artificial intelligence," in Proceedings of Annual Meeting of

[32] L. Feinstein, D. Schnackenberg, R. Balupari, and D. Kindred, "Statistical approaches to ddos attack detection and response," in Proceedings DARPA information survivability conference and exposition, vol. 1. IEEE, 2003, pp. 303–314.

[33] Jianfeng Gao, Michel Galley, and Lihong Li. Neural approaches to conversational AI. Foundations and Trends in Information Retrieval, 13(2-3):127–298, 2019.